\documentclass[twocolumn,showpacs,aps,floatfix,prd,preprintnumbers]{revtex4}
\usepackage{epstopdf}
\usepackage{capt-of}
\usepackage{graphicx}
\usepackage{dcolumn}
\usepackage{bm}

\begin{document}

\title{Charged wormholes in $f(R,T)$ extended theory of gravity}

\author{P.H.R.S. Moraes$^{1}$, W. de Paula$^{1}$, R.A.C. Correa$^{1,2}$}
\affiliation{$^{1}$ITA - Instituto Tecnol\'ogico de Aeron\'autica - Departamento de
F\'isica, 12228-900, S\~ao Jos\'e dos Campos, S\~ao Paulo, Brazil}
\affiliation{$^{2}$Scuola Internazionale Superiore di Studi Avanzati (SISSA), via Bonomea, 265, I-34136 Trieste, Italy}

\begin{abstract}
Wormholes are a solution for General Relativity field equations which characterize a passage or a tunnel that connects two different regions of space-time and is filled by some sort of exotic matter, that does not satisfy the energy conditions. On the other hand, it is known that in extended theories of gravity, the extra degrees of freedom once provided may allow the energy conditions to be obeyed and, consequently, the matter content of the wormhole to be non-exotic. In this work, we obtain, as a novelty in the literature, solutions for charged wormholes in the $f(R,T)$ extended theory of gravity. We show that the presence of charge in these objects may be a possibility to respect some stability conditions for their metric. Also,  remarkably, the energy conditions are respected in the present approach.

\end{abstract}

\maketitle

\input epsf.tex 





\section{Introduction}

Wormholes (WHs) are passages through space-time which connect two different
regions of the Universe. It is believed that such objects might be formed in
regions of intense gravitational fields, in which the highly warped
space-time manifold might allow for the existence of such an unexpected
topology.

In 1935, the idea of WH was introduced by Einstein and Rosen \cite%
{einstein/1935}. They investigated WH solutions which contained event
horizons. Such objects are today referred to as Einstein-Rosen bridges.

The subject was revived in 1988 in the paper by Morris and Thorne (MT) \cite%
{morris/1988}. MT firstly realized that WHs can be traversable, which means
that an eventual travel through the WH could be safe enough, and the minimum
requirement for this is that, departing from black holes, traversable WHs must not have event horizons.

The geometrical content of WHs obeying these features is described by the MT
metric, which in Schwarzschild coordinates and natural units (which shall be
adopted throughout the text), reads \cite{morris/1988}

\begin{equation}  \label{i1}
ds^{2}=e^{2\varphi(r)}dt^{2}-\frac{dr^{2}}{1-\frac{b(r)}{r}}%
-r^{2}(d\theta^{2}+\sin^{2}\theta d\phi^{2}).
\end{equation}
In (\ref{i1}), $\varphi(r)$ is the redshift function and $b(r)$ is the shape
function. There are some properties that must be satisfied by these
quantities in order to have traversable WHs, which we are going to visit
later. For now, we quote that the absence of horizons is achieved if $%
\varphi(r)$ is finite everywhere \cite{visser/1995}.

Within General Theory of Relativity (GR) formalism, it is well known that in
the vicinity of the WH, the null energy condition (NEC),

\begin{equation}  \label{i2}
T_{\mu\nu}k^{\mu}k^{\nu}\geq0,
\end{equation}
is violated \cite{morris/1988,visser/1995}. In (\ref{i2}), $T_{\mu\nu}$ is
the energy-momentum tensor and $k^{\mu}$ is a null vector. Such a NEC
violation implies that the weak, strong and dominant energy conditions,
which are going to be presented later, are violated in the same region
above. In this way, WHs are expected to be filled by some kind of exotic
matter.

Such an important issue can be evaded when dealing with WHs in
extended theories of gravity. One can argue that GR is a particular case of
a more fundamental theory of gravity, which contains geometrical and/or
material corrections respectively to the Einstein and energy-momentum
tensors. Although in the solar system regime those correction terms could be
neglected, in galactic and cosmological length scales and in the strong
gravitational field regime, where there are some indications that GR breaks
down, they could rise as plausible solutions to match theory and
observation. Some enlightening works on extended theories of gravity, their importance and applicability can be appreciated in \cite{de_felice/2010,capozziello/2011,capozziello/2008,koyama/2016,carroll/2005,olmo/2011,ishak/2006}.

The importance of the applicability of extended theories of gravity in the
WHs analysis is that the correction terms may allow the material content
inside these objects to respect the energy conditions. Such a feature was
obtained for different theories of gravitation, such as $f(R)$ and
Gauss-Bonnet theories \cite{mazharimousavi/2016,maeda/2008}.

In the present paper, we will be particularly concerned with charged WHs
(CWHs). The inception of charge in WHs was considered as a possibility to
respect some stability conditions for the metric of these objects in \cite%
{kim/2001}, although the referred authors were not concerned with the
respectability of the energy conditions. CWHs were studied in teleparallel
gravity with non-commutative background in \cite{sharif/2014}. They
concluded that only exotic matter is capable of forming WHs in this
formalism. Furthermore, CWHs were shown to be unstable in ghost scalar field
models and scalar-tensor theories of gravity \cite%
{gonzalez/2009,bronnikov/2005}.

Our aim in the present article is to obtain CWH solutions that respect the
energy conditions in a particular extended theory of gravity, named $f(R,T)$
gravity \cite{harko/2011}. The $f(R,T)$ gravity considers general terms in $%
R $ and $T$ in its gravitational action, for which $R$ stands for the Ricci
scalar while $T$ is the trace of the energy-momentum tensor. The
consideration of the terms proportional to $T$ is motivated by the possible
existence of imperfect fluids in the Universe. Since the matter
content of WHs is described by an imperfect anisotropic fluid, the
consideration of these objects in such a theory is well
motivated. In fact, the analysis of non-charged WHs in $f(R,T)$ gravity has
already been accomplished \cite{yousaf/2017,ms/2017,mcl/2017,zubair/2016,azizi/2013}. Nevertheless, the $f(R,T)$ theory of gravity has been applied to other cosmological and astrophysical purposes \cite{das/2016,alhamzawi/2016,zaregonbadi/2016,amam/2016,ms/2017b,das/2017}.

\section{Charged wormhole metric}

\label{sec:cwhm}

The CWH metric reads \cite{kim/2001}

\begin{equation}  \label{cwhm1}
ds^{2}=\left[e^{2\varphi(r)}+\frac{q^{2}}{r^{2}}\right]dt^{2}-\frac{dr^{2}}{%
1-\frac{b(r)}{r}+\frac{q^{2}}{r^{2}}}-r^{2}(d\theta^{2}+\sin^{2}\theta
d\phi^{2}).
\end{equation}
In Eq.(\ref{cwhm1}), $q$ is the electric charge and $b(r)$ must obey the
following conditions \cite{visser/1995}: i) $b(r_0)=r_0$; ii) $b^{\prime
}(r_0)\leq1$; iii) $b^{\prime }(r)<b(r)/r$, with $r_0$ being the CWH throat
radius and primes indicating radial derivatives. Moreover, in \cite{kim/2001}%
, $\varphi(r)$ was taken as null.

\section{The $f(R,T)$ gravity formalism}\label{sec:frt}

The $f(R,T)$ gravity starts from the action \cite{harko/2011}

\begin{equation}  \label{frt1}
S=\frac{1}{16\pi}\int d^{4}x\sqrt{-g}f(R,T)+\int d^{4}x\sqrt{-g}L.
\end{equation}
In (\ref{frt1}), $g$ is the determinant of the metric, $f(R,T)$ is the
function of $R$ and $T$ and $L$ is the lagrangian density.

If $L$ depends only on the metric components and not on its derivatives,
one has, for the energy-momentum tensor, the following:

\begin{equation}  \label{frt2}
T_{\mu\nu}=g_{\mu\nu}L-2\frac{\partial L}{\partial g^{\mu\nu}},
\end{equation}
so that $T=g^{\mu\nu}T_{\mu\nu}$.

By varying action (\ref{frt1}) with respect to $g_{\mu\nu}$ for the case $%
f(R,T)=R+2\lambda T$, with constant $\lambda$, one obtains the following field equations

\begin{equation}  \label{frt3}
G_{\mu\nu}=8\pi T_{\mu\nu}^{eff},
\end{equation}
with $G_{\mu\nu}$ being the Einstein tensor and the effective
energy-momentum tensor

\begin{equation}  \label{frt4}
T_{\mu\nu}^{eff}=T_{\mu\nu}-\frac{\lambda}{4\pi}\left(T_{\mu\nu}+\Theta_{\mu%
\nu}-\frac{1}{2}Tg_{\mu\nu}\right),
\end{equation}
with

\begin{equation}  \label{frt5}
\Theta_{\mu\nu}\equiv g^{\alpha\beta}\frac{\partial T_{\alpha\beta}}{%
\partial g^{\mu\nu}}.
\end{equation}
Such a functional form for $f(R,T)$ is the simplest case with material corrections. It was proposed by the authors of the theory in \cite{harko/2011} and it has been deeply applied since them as one can check, for instance, in \cite{ms/2017,mcl/2017,azizi/2013,das/2016,das/2017}. 

In the present case, we can write

\begin{equation}  \label{frt6}
G_{\mu\nu}=8\pi[T_{\mu\nu}^{eff(m)}+T_{\mu\nu}^{eff(e)}],
\end{equation}
with $T_{\mu\nu}^{eff(m)}$ being the effective energy-momentum of matter
threading the WH and $T_{\mu\nu}^{eff(e)}$ the effective electromagnetic
energy-momentum tensor.

T. Harko et al. in \cite{harko/2011} have constructed $\Theta_{\mu\nu}^{(m)}$
and $\Theta_{\mu\nu}^{(e)}$ respectively from

\begin{equation}  \label{frt7}
L^{(m)}=-\mathcal{P}
\end{equation}
and

\begin{equation}  \label{frt8}
L^{(e)}=-\frac{1}{16\pi}F_{\alpha\beta}F_{\gamma\sigma}g^{\alpha\gamma}g^{%
\beta\sigma},
\end{equation}
with $\mathcal{P}$ being the fluid total pressure and $F_{\alpha\beta}$ the
electromagnetic field tensor, and found

\begin{equation}  \label{frt9}
\Theta_{\mu\nu}^{(m)}=-2T_{\mu\nu}^{(m)}-pg_{\mu\nu}
\end{equation}
and

\begin{equation}  \label{frt10}
\Theta_{\mu\nu}^{(e)}=-T_{\mu\nu}^{(e)}.
\end{equation}

It is worth to quickly discuss the choice in Eq.(\ref{frt7}). Gravity theories that allow the geometrical and material sectors to be coupled (besides $f(R,T)$ models, check also $f(R,L_m)$ theories \cite{harko/2010}, with $L_m$ being the matter lagrangian) predict the movement of test particles in gravitational fields to be non-geodesic. Rather, according to these theories, the particles move in the presence of an extra force. It has been shown that when considering the lagrangian to be proportional to $\mathcal{P}$, as in Eq.(\ref{frt7}), such an extra force vanishes \cite{bertolami/2008,harko/2014}.

In this manner, from (\ref{frt9})-(\ref{frt10}), we can rewrite our field equations as

\begin{equation}  \label{frt11}
G_{\mu\nu}=8\pi\left[T_{\mu\nu}^{(m)}+T_{\mu\nu}^{(e)}+\frac{\lambda}{4\pi}%
\Pi_{\mu\nu}\right],
\end{equation}
in which we have defined the tensor

\begin{equation}  \label{frt12}
\Pi_{\mu\nu}\equiv T_{\mu\nu}^{(m)}+\mathcal{P}g_{\mu\nu}+\frac{1}{2}%
[T^{(m)}+T^{(e)}]g_{\mu\nu},
\end{equation}
with $T^{(m)}=g^{\mu\nu}T_{\mu\nu}^{(m)}$ and $T^{(e)}=g^{\mu\nu}T_{\mu%
\nu}^{(e)}$.

\section{Charged wormholes in $f(R,T)$ gravity}

\label{sec:cwhfrt}

We assume that an anisotropic fluid of energy-momentum tensor

\begin{equation}  \label{cwhfrt0a}
T_{\mu\nu}^{(m)} = \text{diag} (\rho,-p_r,-p_t,-p_t)
\end{equation}
is filling in the WH, with $\rho$ being the matter-energy density and $p_r$
and $p_t$ the radial and transverse components of the pressure, such that $%
\mathcal{P}=(p_r+2p_t)/3$.

We also take only the radial component of the electric field in $%
F_{\alpha\beta}$, so that for the metric (\ref{cwhm1}), the energy-momentum tensor for the electromagnetic field is

\begin{equation}
T_{\mu \nu }^{(e)}=\frac{1}{8\pi } E^{2} \, \text{diag}(3,3,1,1) \, \gamma \, \kappa ,
\label{cwhfrt0b}
\end{equation}%
in which $E=E(r)=(q/r^{2})\sqrt{\left\vert g_{00}g_{11}\right\vert }$ is the
radial component of the electric field and we defined
\begin{eqnarray}  \label{cwhfrt0c}
\gamma &\equiv & \frac{1}{q^{2}+e^{2\varphi}r^{2}} , \nonumber\\
\kappa &\equiv & q^{2}+r^{2}-br \, . 
\end{eqnarray}

Therefore, by following the approach of the previous sections, one can write the field equations (\ref{frt11}) as:

\begin{equation}
b^{\prime }=\left( 2+\frac{\lambda }{\pi }\right) \frac{q^{2}}{r^{2}}+(8\pi
+3\lambda )r^{2}\rho -\frac{\lambda \left( p_{r}+8p_{t}\right) r^{2}}{3}.
\label{s1}
\end{equation}%
\begin{eqnarray}
&&\left. \frac{\gamma }{r^{2}}\left[ \frac{q^{2}}{r^{2}}(\kappa
+r^{2})-e^{2\varphi }(q^{2}-rb+2r\kappa \varphi ^{\prime })\right] =\right. 
\nonumber \\
&&\left. -8\pi p_{r}+3E^{2}\gamma \kappa +\lambda \left( \rho -\frac{%
7p_{r}-2p_{t}}{3}+\frac{E^{2}\gamma \kappa }{\pi }\right) ,\right. \nonumber\\
\label{t1}
\end{eqnarray}%
\begin{eqnarray}
&&\left. \frac{e^{2\varphi }}{q^{2}}\{(r\varphi ^{\prime }+1)[r\left(
-rb^{\prime }+b+2\kappa \varphi ^{\prime 2}\right) +2\kappa r^{2}\varphi
^{\prime \prime }]\}\right.   \nonumber \\
&&\left. +\frac{1}{r^{4}}[kq^{2}+r^{2}e^{2\varphi }(2q^{2}r\varphi ^{\prime
2}+2r^{2})]-b^{\prime }(r\varphi ^{\prime }+1)\right.   \nonumber \\
&&\left. +2[(q^{2}+r^{2})(\varphi ^{\prime \prime }+2\varphi ^{\prime
2})+3r\varphi ^{\prime }]\right.   \nonumber \\
&&\left. -b\left[ 2r\varphi ^{\prime \prime }+\varphi ^{\prime }(4r\varphi
^{\prime }+5)+\frac{3}{r}\right] \right.   \nonumber \\
&&\left. =\frac{2e^{-2\varphi }}{\gamma ^{2}q^{2}}\left[ p_{t}-E^{2}\gamma
\kappa -\lambda \left( \rho -\frac{p_{r}+8p_{t}}{3}+\frac{E^{2}\gamma \kappa 
}{\pi }\right) \right] .\right. \nonumber\\
\label{rho1}
\end{eqnarray}

Our next goal is to find analytical solutions for the coupled system of equations (\ref{s1})-(\ref{rho1}).

\section{Analytical solutions}

In this section, we will show a class of analytical solutions for the present CWH.  We will obtain the solutions by analyzing one equation at a time, i.e., from Eq.(\ref{s1}), we will fix the parameters $\alpha$ and $\beta$ and solve for $b(r)$. Then we find $\varphi (r)$ that solves Eq.(\ref{t1}). Finally, we present the density $\rho(r)$ by solving Eq.(\ref{rho1}). 

Let us write the equation of state (EoS) for the matter filling in the concerned CWH to be

\begin{equation}
p_{r}=\alpha \, \rho ~~ , ~~ p_{t}=\beta \, \rho ,\text{\ }  \label{s2}
\end{equation}
for $\alpha$ and $\beta$ constants. This form for the EoS has been constantly used in the literature, as one can check, for instance, Refs.\cite{ms/2017,rahaman/2007,garcia/2010,bahamonde/2016,shaikh/2015}. In particular, we will choose both constants as functions of the parameter $\lambda$ as

\begin{eqnarray}
\alpha &=&\frac{3}{2}\left( \frac{\lambda +2\pi }{\lambda +3\pi }\right) , \\
\beta &=&\frac{3}{4}\left[ 5+\pi \left( \frac{16}{\lambda }+\frac{1}{\lambda
+3\pi }\right) \right],
\end{eqnarray}
such that, by replacing them in Eq.(\ref{s1}) we find 

\begin{equation}
b(r)=-\frac{\Omega _{0}}{r}  \label{s5}
\end{equation}
as our solution for the shape function, where $\Omega _{0}\equiv (2+\lambda /\pi )$. In order to make $b(r)$ to satisfy the metric conditions presented in Section \ref{sec:cwhm}, it is necessary to impose that $\lambda <-2\pi $. 

Now, by substituting Eq.(\ref{s5}) into (\ref{t1}), we can obtain the corresponding solution for the redshift function%

\begin{eqnarray}
\varphi (r) &=&\ln \left\{ \frac{c}{\sqrt{r}(Q_{0}+r^{2})^{\lambda _{0}}}%
\right.  \nonumber \\
&&\left.+\frac{Q_{1}}{(Q_{0}+r^{2})^{\lambda _{0}}}\left[
2\, Q_{0}^{\lambda _{0}-1}(Q_{0}+r^{2})^{\lambda _{0}}+\right. \right. 
\nonumber \\
&&\left. \left. Q_{2} \, F\left( \frac{1}{2},\frac{2-\lambda }{4},\frac{5}{4};-%
\frac{r^{2}}{Q_{0}}\right) \right] \right\} ,  \label{s12}
\end{eqnarray}

\noindent where $c$ is an arbitrary constant of integration and $F(1/2,$ $%
(2-\lambda )/4,5/4;-r^{2}/Q_{0})$ is the so-called hypergeometric
function. Furthermore, we use the
following definitions

\begin{eqnarray}
Q_{0} &\equiv &q^{2}+\Omega _{0}, \\
Q_{1} &\equiv &\frac{1}{(\lambda +3) Q_{0}}, \\
Q_{2} &\equiv &(\lambda -1)\Omega _{0}-q^{2}\left[ 8+\lambda (\lambda +5)%
\right] , \\
\lambda _{0} &\equiv &(\lambda +2)/4.
\end{eqnarray}

\section{Energy conditions}\label{sec:ec}

To complete our objective of solving the field equations for a CWH in the $f(R,T)$ model under analysis, we have to find the density $\rho$ that fullfills  Eq.(\ref{rho1}). 

Indeed we find a solution for the density in terms of the parameters of the model. To illustrate the density behavior we present Fig.\ref{rho} below, in which we show the 
dependence of the density on $r$ and some model parameters, as the CWH charge. In particular, we obtain that by increasing the charge, it is necessary to 
have a higher mass-energy density in order to maintain the WH geometry.

\begin{figure}[]
\includegraphics[scale=0.35]{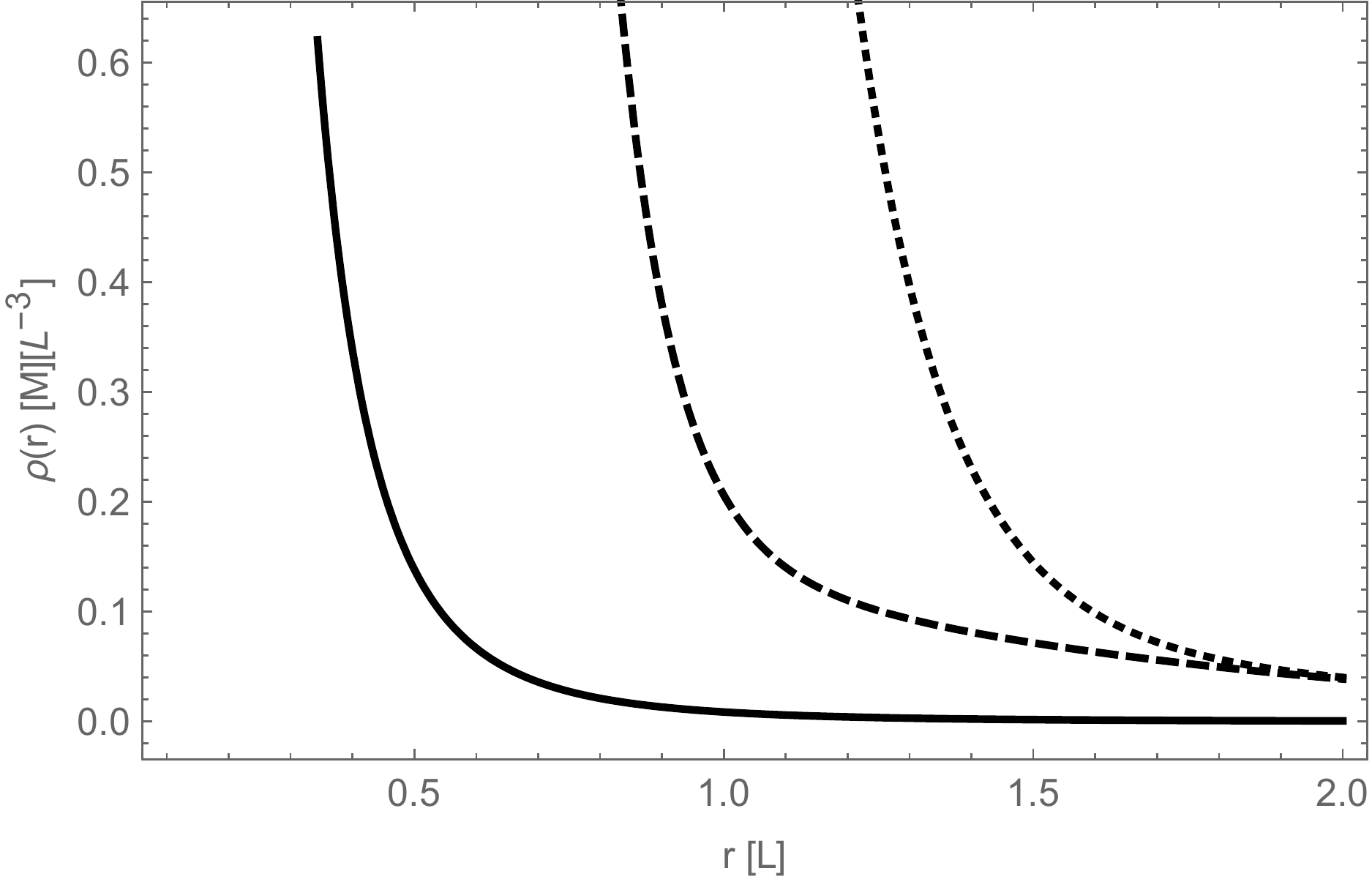}
\caption{Mass-energy density as a function of the radial variable for the CWH charge given by $q=0.5$ (thin line), $q=2.5$ (dashed line) and $q=4.5$ (dotted line). We consider, for all curves, $\lambda=-7$ and $c=153$.}
\label{rho}
\end{figure}

We have now to determine the region of parameters where our solutions satisfy the energy conditions \cite{morris/1988,visser/1995}. 

The weak energy condition can be written as 

\begin{eqnarray}
&&\rho + p_t \ge 0 \label{WEC1}\, ,\\
&&\rho + p_r \ge 0 \label{WEC2}\, ,
\end{eqnarray}
\noindent while the strong energy condition reads

\begin{eqnarray}
&&\rho + p_t + p_r  \ge 0 \, .\label{SEC}
\end{eqnarray}

\begin{figure}[]
\includegraphics[scale=0.22]{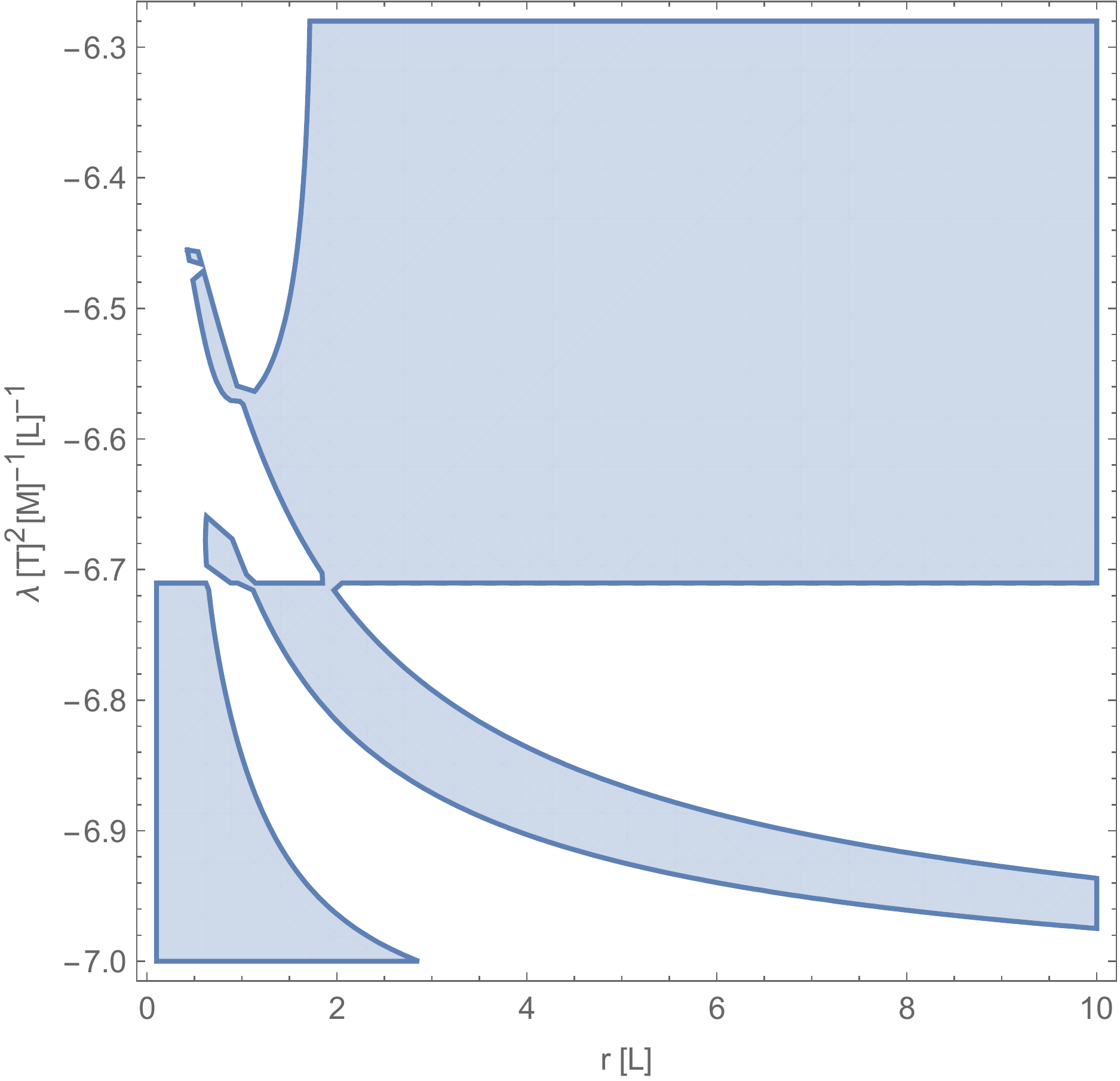} \includegraphics[scale=0.22]{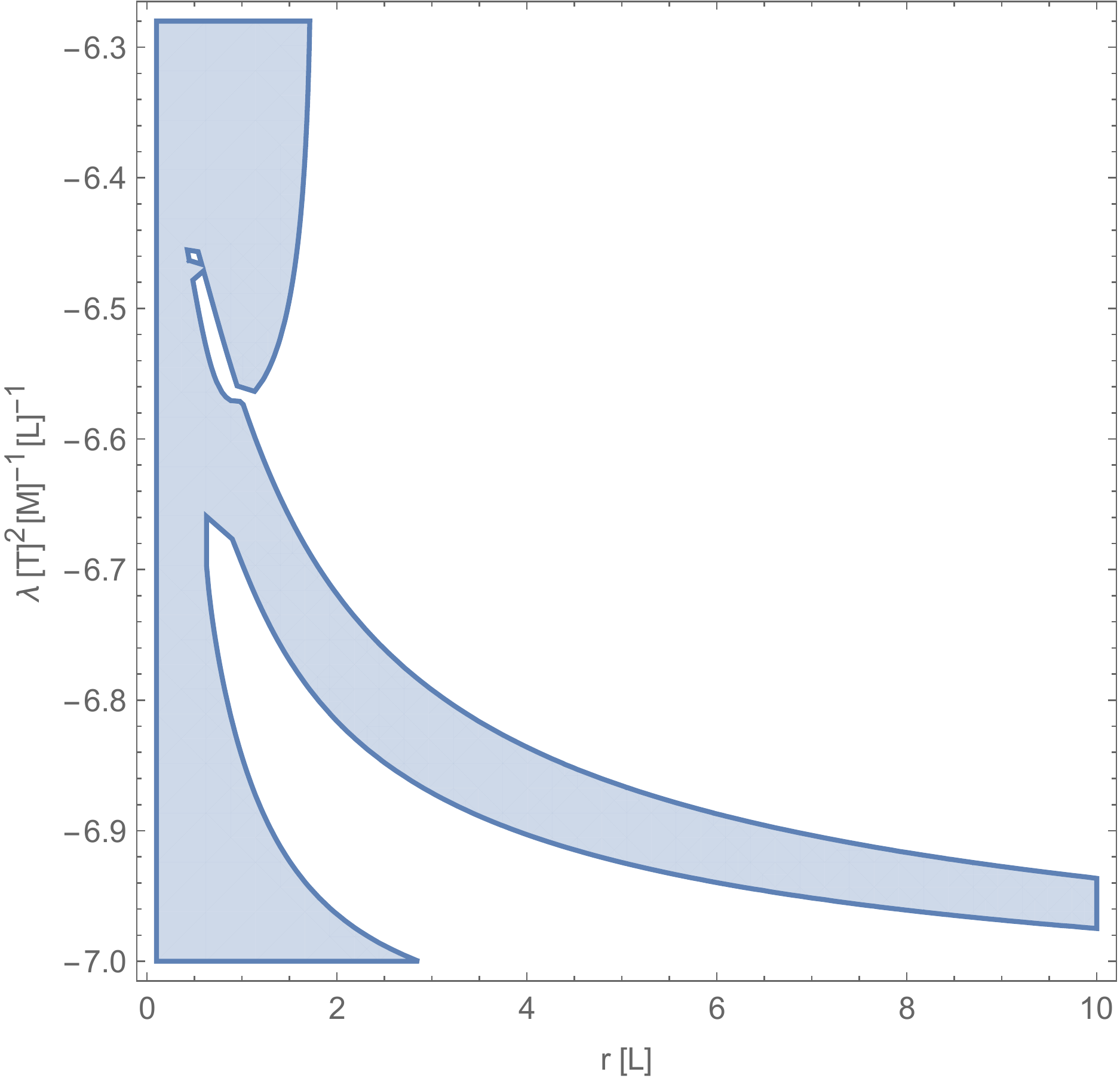}
\caption{Region of the parameter $\lambda$ that satisfies the weak energy condition as a function of the radial variable. In the left panel it is represented the constrain given by Eq.(\ref{WEC1}) and in the right panel, the one given by Eq.(\ref{WEC2}). The CWH electric charge is $q = 0.5$ and $c=153$.}
\label{WECfig}
\end{figure}

\begin{figure}[h!]
\includegraphics[scale=0.25]{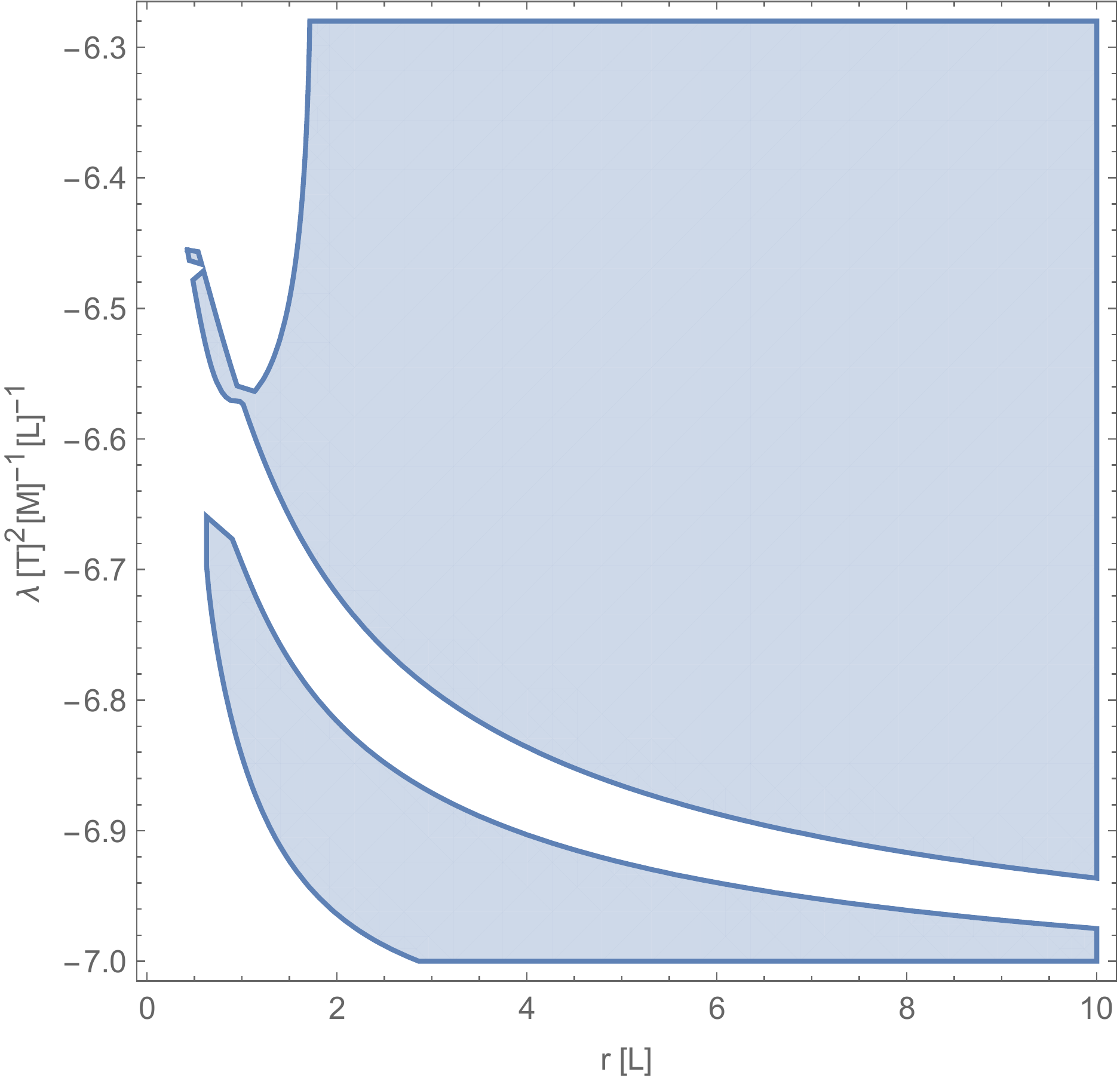}
\caption{Region of the parameter $\lambda$ that satisfies the strong energy condition constrain as a function of the radial variable. The CWH electric charge is $q = 0.5$ and $c=153$ .}
\label{SECfig}
\end{figure}

The inequalities expressed by Eqs.(\ref{WEC1})-(\ref{SEC}) define regions of validity for the parameter $\lambda$, which was already constrained to $\lambda <-2\pi $ by the metric conditions discussed, earlier, in Section \ref{sec:cwhm}. In Figure  \ref{WECfig} we show the region of parameters that satisfies the weak conditions and 
in Figure \ref{SECfig}, the region of parameters that satisfies the strong energy constrain. We recall that we are working with natural units and therefore we express the radial coordinate as dimension of length [L] and $\lambda$ as a combination of the dimension of time [T], mass [M] and length [L].

In Fig.\ref{SECfig}, we show the strong energy condition application. We highlight the regions of the parameter $\lambda$ that satisfy the strong energy condition in terms of the radial coordinate for a fixed charge.  

The Figures \ref{WECfig} and \ref{SECfig} show that we succeeded to find a family of solutions for the CWH-$f(R,T)$ gravity field equations that fulfill the weak and strong energy conditions criteria. Such an achievement is not possible to be attained using GR.

\section{Conclusions}\label{sec:con}

WHs have been proposed in the literature by MT as a tool for teaching GR \cite{morris/1988}. Over the years, the interest in these objects has grown and today there are several proposals to detect them, with the most popular of them being based on the gravitational (micro)lensing \cite{kuhfittig/2014,abe/2010,toki/2011}. Another well motivated possibility of detection may come from the distinction between WHs and black holes \cite{tsukamoto/2012,li/2014}.

In this article, we have constructed, for the first time in the literature, CWHs in the $f(R,T)$ theory of gravity. The modelling of WHs in extended theories of gravity is motivated by the possibility of these objects to be filled by non-exotic matter, that is, matter that respects the energy conditions. The $T$-dependence suggested in $f(R,T)$ gravity comes from the possible existence of imperfect fluids in the universe. In this way, the WH analysis in such a theory is well motivated, since the matter content of these objects is described by an anisotropic imperfect fluid.

We have worked with a functional form for $f(R,T)$ that presents correction terms only in the material - and not in the geometrical - sector of the theory. By following the approach presented in \cite{harko/2011}, we have considered both (\ref{frt7}) and (\ref{frt8}) as lagrangians for our system and those led us to the field equations (\ref{frt11}). 

Let us recall our solution (\ref{s5}) for the shape function. The $r$-proportionality in (\ref{s5}), that is, $r^{-1}$, is the same used by Kim and Lee in Ref.\cite{kim/2001}, in which the metric of a CWH was firstly introduced. The same form for $b(r)$ was used in $f(R,T)$ and $f(R)$ theories, as it can be checked, respectively, in \cite{zubair/2016,lobo/2009} and for WHs with a cosmological constant \cite{lemos/2003}. We remark that while such a form for $b(r)$ was assumed {\it a priori} in these references, in the present article it has been obtained as a model solution.

Furthermore, it can be seen that the solution for the redshift function (\ref{s12}) is finite everywhere, as request in order to have traversable WHs \cite{morris/1988,visser/1995}.

To finish, we have shown that, remarkably, the CWH under analysis respects the weak and strong energy conditions for a wide range of values for the parameter $\lambda$ (check Figs.\ref{WECfig}-\ref{SECfig}).

\

\acknowledgments PHRSM thanks S\~ao Paulo Research Foundation
(FAPESP), grant 2015/08476-0, for financial support. WP thanks CAPES and CNPq for financial support.
RACC would like to thank FAPESP, grants 2016/03276-5 and 2017/26646-5, for financial support.

\end{document}